\documentstyle[aps,a4]{revtex}

\input{psfig}
\psrotatefirst

\begin{document}

\title{\boldmath Muon Catalyzed Fusion in 3~K Solid Deuterium.}

\author{P.E.~Knowles\footnote{Present Address: Institut de Physique Nucl\'eaire,
         Universit\'e Catholique de Louvain, B--1348,
         Louvain--La--Neuve, Belgium.},
        G.A.~Beer, G.R.~Mason, T.A.~Porcelli}
  \address{University of Victoria, Victoria B.C., Canada, V8W 2Y2}
\author{A.~Adamczak}
  \address{Institute of Nuclear Physics, Cracow, Poland}
\author{J.M.~Bailey} 
  \address{Chester Technology, England}
\author{J.L.~Beveridge, G.M.~Marshall and A~Olin}
  \address{TRIUMF, 4004 Wesbrook Mall, Vancouver B.C., Canada, V6T 2A3}
\author{M.C.~Fujiwara}
  \address{University of British Columbia, Vancouver B.C., Canada, V6T 2A6}
\author{T.M.~Huber}
  \address{Gustavus Adolphus College, St. Peter, Minnesota 56082}
\author{R.~Jacot-Guillarmod, and F.~Mulhauser}
  \address{Institut de Physique, Universit\'e de Fribourg, CH-1700 Fribourg, Switzerland}
\author{P.~Kammel\footnote{Present Address: University of California,
        LBNL, Berkeley, California 94720}, and J.~Zmeskal}
  \address{Austrian Academy of Sciences, A--1090 Vienna, Austria}
\author{S.K.~Kim}
  \address{Jeonbuk National University, Jeonju City 560--756, South~Korea}
\author{A.R.~Kunselman}
  \address{University of Wyoming, Laramie, Wyoming 82071}
\author{C.J.~Martoff}
  \address{Temple University, Philadelphia, Pennsylvania 19122}
\author{C.~Petitjean}
  \address{Paul Scherrer Institut, PSI, CH--5232 Villigen, Switzerland}

\maketitle

\begin{abstract}

Muon catalyzed fusion in deuterium has traditionally been studied in
gaseous and liquid targets. The TRIUMF solid-hydrogen-layer target
system has been used to study the fusion reaction rates in the solid
phase of D$_2$ at a target temperature of 3~K\@.  Products of two
distinct branches of the reaction were observed; neutrons by a liquid
organic scintillator, and protons by a silicon detector located inside
the target system. The effective molecular formation rate from the
upper hyperfine state of $\mu d$ and the hyperfine transition rate
have been measured:
$\tilde{\lambda}_{\frac{3}{2}}=2.71(7)_{stat.}(32)_{syst.}\:\mu
\mathrm{s}^{-1}$, and $\tilde{\lambda}_{\frac{3}{2}\frac{1}{2}}
=34.2(8)_{stat.}(1)_{syst.}\:\mu \mathrm{s}^{-1}$. The molecular
formation rate is consistent with other recent measurements, but not
with the theory for isolated molecules. The discrepancy may be due to
incomplete thermalization, an effect which was investigated by Monte
Carlo calculations. Information on branching ratio parameters for the
s~and p~wave $d+d$ nuclear interaction has been extracted.

\end{abstract}

\pacs{36.10.Dr, 36.10.-k}%

\newpage

\section{Introduction}

For forty years it has been known that the introduction of a negative
muon into a mixture of the three isotopes of hydrogen (protium,
deuterium, and tritium) can lead to fusion reactions between the
hydrogen nuclei, a process called muon catalyzed fusion or
$\mu{\mathrm{CF}}$. The intricately connected molecular, atomic, and
nuclear processes that occur have taken many years to identify and
offer a richness of physics in both theoretical and experimental
domains \cite{breun89,ponom90,froel92,cohen93}.

The TRIUMF solid-hydrogen-layer target system \cite{knowl96} was
used to measure $\mu{\mathrm{CF}}$ in solid deuterium for the first
time. The preliminary results \cite{marsh93,kamme93} from our
experiment at low temperature were in sharp disagreement with theory
and initiated further investigations of the reactions in solid D$_2$
\cite{knowl96b,knowlphd,strasphd,deminDUBNA}. This work presents our
final results based on larger statistics and the simultaneous
observation of two distinct branches of the fusion reaction.

The $\mu{\mathrm{CF}}$ reaction in pure deuterium has been well
investigated for gaseous and liquid targets---see
Refs.~\cite{scrin93,zmesk90} and references therein---where the
assumption that the D$_2$ is not interacting with its neighbors has
been used in the theoretical analysis. That approximation has led to
generally good agreement between theory and experiment when the
deuterium is a fluid. The data presented here from $\mu{\mathrm{CF}}$
in solid deuterium maintained at 3~K indicate that it fails. 

\section{Theory}
\subsection{Physical Processes}

A negative muon introduced into pure deuterium slows and captures on a
deuteron to form an excited $\mu d$ in a characteristic time of
picoseconds in high density targets such as solids or liquids
\cite{cohen83,koren76,leonx62}. The subsequent deexcitation of the
muonic atom occurs via Stark, Auger, scattering, radiative, and
transfer processes, which occur on the 100~ps time scale
\cite{czapl94}. The process $(\mu d)_{n_i}+d\rightarrow(\mu
d)_{n_f}+d$ with $n_f < n_i$, often called Coulomb deexcitation,
provides a source of acceleration for the muonic atom since the final
state contains only the massive particles which share the energy
released in the $n_i\rightarrow n_f$ transition.  The subsequent
energy distribution of the $\mu d$ contains both thermal and
epithermal (nonthermalized) components of muonic atoms. The relative
populations reflect the temperature and density of the muonic atoms'
environment \cite{asche95,marku94,kraim89,abbotsub}.

Once the atomic ground state has been reached, the reactions of the
$\mu d$ are dominated by the scattering from D$_2$ molecules,
including inelastic processes such as the realignment of the $\mu d$
hyperfine state, or the formation of the molecular bound state $d\mu d$.

A muonic deuterium atom may form a molecule or molecular ion via two
mechanisms. The first is the capture of the muonic atom on one nucleus
of a deuterium molecule to form a molecular bound state with the
subsequent release of the bound state energy by an Auger process:
\begin{equation}
 \mu d + [ddee] \rightarrow [(d\mu d)de] +  e_{\mathrm Auger}.
\label{equ:augerform}
\end{equation}

A second process, a resonance mechanism first proposed by Vesman
\cite{vesma67}, is a curiously fortuitous effect that depends on the
energy levels of all the involved bound states. The existence of a
state of the $( d\mu d)$ bound by $\sim$2~eV allows the excess energy
of the system, the energy liberated in molecular formation plus the
incident $\mu d$ energy, to be absorbed in the excitation of
rotational and vibrational states, $(\nu_{f}K_{f})$, of the full six
body final state (strongly bound states liberate energy above the
dissociation threshold of the D$_2$). A resonance in the formation
cross section occurs when the incident energy of the $\mu {d}_{F}$ in
hyperfine state~F is such that the total initial energy of the system
matches the energy of some set of final state excitation parameters,
\begin{equation} 
\mu{d}_{F} + [ddee]_{\nu_{i},K_{i}} \rightarrow
\left [ \left(d\mu{d}\right)_{J\nu}{dee}\right]_{\nu_{f},K_{f}}.
\label{equ:resform}
\end{equation}
Here, $J$ and $\nu$ refer to the orbital angular momentum and
vibrational quantum number respectively, while the $\nu_{i,f}$,and
$K_{i,f}$ refer to the vibrational quantum number and rotational
quantum number of the six-body system. The resonant formation process
has been well studied, resulting in calculations of the energy
dependence of the rate
\cite{mensh87,faifm89,faifm89b,scrin90,faifmDUBNA}.

\begin{figure}
\centerline{
\setlength{\unitlength}{0.635cm}
\begin{picture}(14,10)(0,0)
\put(0,0){\psfig{figure=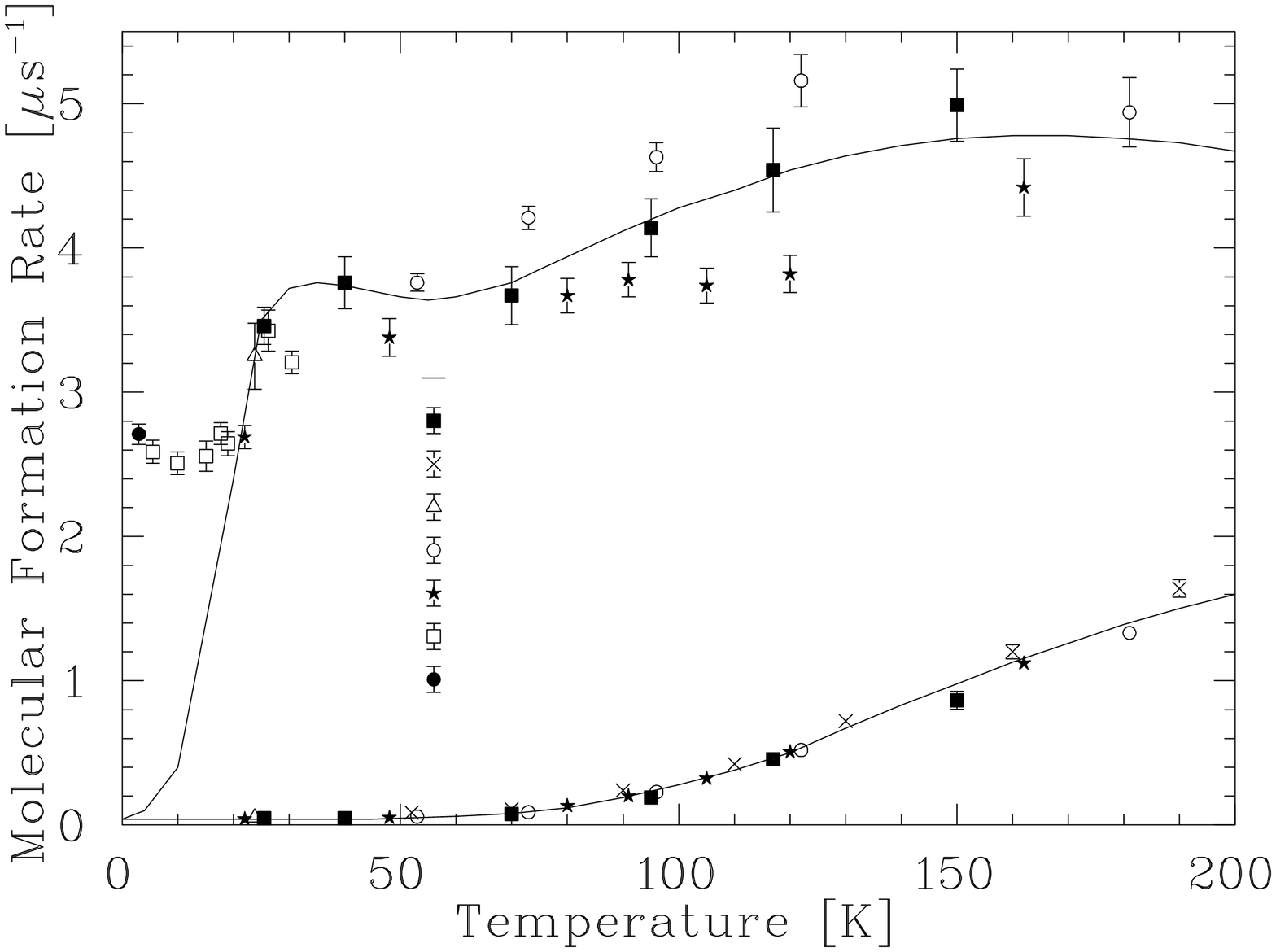,width=8.89cm,angle=90}}
\put(2.8,2.1){\mbox{$\tilde{\lambda}_{\frac{1}{2}}(T)$}}
\put(2.8,8.5){\mbox{$\tilde{\lambda}_{\frac{3}{2}}(T)$}}
\put(4.9,6.2){\makebox(0,0)[l]{\footnotesize Theory: Scrinzi {\it et~al.}  \cite{scrin93}}}
\put(4.9,5.7345){\makebox(0,0)[l]{\footnotesize Gas: Zmeskal {\it et~al.}  \cite{zmesk90} }}
\put(4.9,5.2691){\makebox(0,0)[l]{\footnotesize Gas: Balin {\it et~al.} \cite{balin92}}}
\put(4.9,4.8036){\makebox(0,0)[l]{\footnotesize Liquid: N\"agele {\it et~al.} \cite{nagel89}}}
\put(4.9,4.3382){\makebox(0,0)[l]{\footnotesize Gas: Petitjean {\it et~al.} \cite{petitDUBNA}}}
\put(4.9,3.8727){\makebox(0,0)[l]{\footnotesize Liquid/Gas: Dzhelepov {\it et~al.} \cite{dzhel92c}}}
\put(4.9,3.4073){\makebox(0,0)[l]{\footnotesize Solid: Demin {\it et~al.} \cite{deminDUBNA} }}
\put(4.9,2.9418){\makebox(0,0)[l]{\footnotesize Solid: This work }}
\end{picture} }
\caption{Molecular formation rates found by the two--node
approximation fit to data measured in solid, liquid and gas, and
calculations of rates in liquid and gas.  Statistical uncertainties
are given, and the $\tilde{\lambda}_{\frac{3}{2}}$ values of Demin
{\it et~al.} \protect\cite{deminDUBNA} have been normalized to
$\tilde{\lambda}_{\frac{1}{2}}=0.044$ over their 5--30~K temperature
range.}
\label{fig:lamdrate}
\end{figure}

Calculation of an effective formation rate which can be
compared to experiment must assume an energy distribution of
the $\mu d$ atoms in the D$_2$ target. The distribution,
which is determined by the initial energy at which the $\mu
d$ was formed and the subsequent energy loss processes, is
convoluted with the reaction rates at an appropriate
temperature to produce an overall temperature dependent rate
\cite{scrin93}. The agreement between theory, which predicts
a rapid thermalization of the $\mu d$ and hence uses a $\mu d$
population in thermal equilibrium with the surrounding deuterium, and
experiments which use liquid and gas targets, is good
(see~Fig.~\ref{fig:lamdrate}). The results of present measurements in
solids are in disagreement.

The small size characteristic of the muonic molecular
bound states is such that nuclear fusion can occur.
For deuterium, there are two fusion channels:
\begin{eqnarray}
d\mu d &\longrightarrow& \mu + t + p + 4.0\mathrm{\:MeV} \nonumber \\ 
       &\longrightarrow& \left \{ \begin{array}{l}
\mu +{}^3{\mathrm{He}} + n + 3.3\mathrm{\:MeV} \\
\mu{}^3{\mathrm{He}} + n + 3.3\mathrm{\:MeV}
                                  \end{array} \right.
\end{eqnarray}
Once fusion occurs, the muon is generally set free and can begin the
process again (cycling); however, the $Z=2$ helium nucleus can
sometimes capture the muon into atomic orbital bound states stopping
further reactions, a process called sticking.  The number of fusion
cycles in which the muon participates is limited by the muon lifetime
and by the sticking probability. The kinetics for the cycle shown in
Fig.~\ref{fig:muind} are given in the Appendix.

\begin{figure}
\centerline{
\setlength{\unitlength}{0.6350cm}
\begin{picture}(14,10)(0,0)
\put(0,0){\psfig{figure=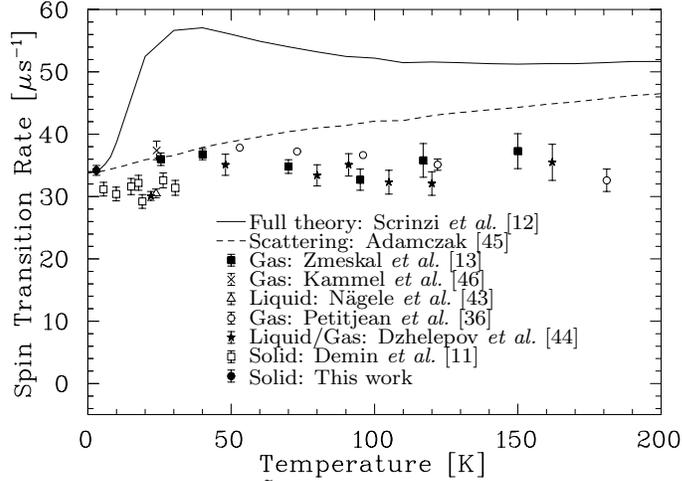,width=8.89cm,angle=90}}
\put(4.99,5.3){\makebox(0,0)[l]{\footnotesize Full theory: Scrinzi {\it et~al.} \cite{scrin93}}}
\put(4.99,4.9){\makebox(0,0)[l]{\footnotesize Scattering: Adamczak \cite{adamc89}}}
\put(4.99,4.5){\makebox(0,0)[l]{\footnotesize Gas: Zmeskal {\it et~al.} \cite{zmesk90} }}
\put(4.99,4.1){\makebox(0,0)[l]{\footnotesize Gas: Kammel {\it et~al.} \cite{kamme82} }}
\put(4.99,3.7){\makebox(0,0)[l]{\footnotesize Liquid: N\"agele {\it et~al.} \cite{nagel89}}}
\put(4.99,3.3){\makebox(0,0)[l]{\footnotesize Gas: Petitjean {\it et~al.} \cite{petitDUBNA}}}
\put(4.99,2.9){\makebox(0,0)[l]{\footnotesize Liquid/Gas: Dzhelepov {\it et~al.} \cite{dzhel92c}}}
\put(4.99,2.5){\makebox(0,0)[l]{\footnotesize Solid: Demin {\it et~al.} \cite{deminDUBNA}}}
\put(4.99,2.1){\makebox(0,0)[l]{\footnotesize Solid: This work}}
\end{picture} }
\caption{Hyperfine transition rate $\tilde{\lambda}_{\frac{3}{2}\frac{1}{2}}$.
Theory calculations are shown for scattering only (dashed line) and
for scattering plus back decay (solid line).}
\label{fig:lamqd}
\end{figure}

The energy difference between the two hyperfine states of the $\mu d$
is 48.5~meV; the energy is readily liberated in processes which
depopulate the upper spin state.  There is a disagreement between
theory and experiment for the hyperfine transition rate
$\tilde{\lambda}_{\frac{3}{2}\frac{1}{2}}$, a situation which is well
discussed in Scrinzi {\it et~al.}~\cite{scrin93}. As indicated in
Fig.~\ref{fig:lamqd}, theory models the shape of the temperature
dependence, but the calculated rate is too high. Given the agreement
between the calculated and measured formation rates
(Fig.~\ref{fig:lamdrate}) and the accuracy in the calculations for the
scattering reactions (which are expected to be correct to 10\%), the
discrepancy is not understood.

\subsection{Monte Carlo Studies}
 The two-node approximation (Appendix) assumes that the rates
used to represent the various processes are time independent, or,
equivalently, the result after convoluting the $\mu d$ energy
distribution with the energy dependence of the rates remains constant
in time.  Since the calculations of the muonic processes give energy
dependent rates, it is natural to use the Monte Carlo method to study
the time evolution of the muonic atom energy distribution and thus the
interplay between all the energy dependent rates.

Until recently, only scattering cross sections for muonic atoms on
free hydrogen isotopes were available. Monte Carlo simulations
predicted that the time required to thermalize a muonic atom after
formation was very short ($20\pm20$~ns in D$_2$ gas at 5\% of liquid
density, shorter times at higher density).  After rapid
thermalization, the energy distribution of the atoms was constant in
time making the formation rates time independent and thus validating
the two-node approximation.

\begin{figure}
\centerline{\psfig{figure=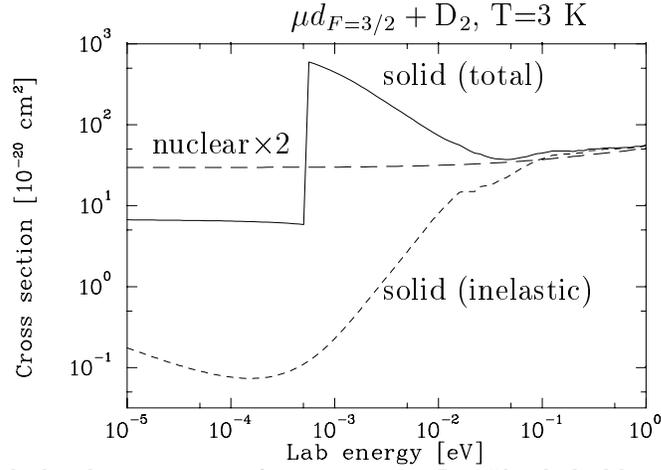,width=8.89cm}}
\caption{Calculated cross sections for scattering in D$_2$.
The dashed line shows the first order approximation to the cross
section on free molecules. Note the logarithmic scale, the suppression
of the inelastic scattering, and the Bragg cutoff for elastic
scattering. See \protect\cite{adamcDUBNA} for a description of the calculations.}
\label{fig:xsol}
\end{figure}

At low energies, $\mu d$ elastic scattering on a solid lattice does
not reduce the $\mu d$ lab kinetic energy due to the large mass of the
recoiling crystal. It was initially suggested by M.~Leon that
mechanisms for energy loss might be curtailed for muonic atoms in
solid hydrogen isotopes.  First quantitative calculations of the
energy dependent scattering cross sections in solids have been
completed \cite{adamcDUBNA}. In Fig.~\ref{fig:xsol} the scattering
cross section for $F=\frac{3}{2}$ $\mu d$ atoms on a solid
ortho-deuterium target at 3~K and zero pressure is shown.  The total
cross section includes elastic scattering on the {\it fcc} lattice of
frozen D$_2$ molecules and inelastic scattering processes---phonon
transitions and rovibrational excitations of the D$_2$ molecules. The
$\mu d$ deceleration rate is determined by the inelastic processes
because of the large recoil mass in elastic scattering processes.  The
rovibrational excitation threshold is about 10~meV, and below that the
$\mu d$ atoms lose energy in purely phonon scattering, an amplitude
which also falls rapidly with decreasing $\mu d$ energy.  In
Fig.~\ref{fig:xsol} the inelastic cross section for $\mu d$ in solid
D$_2$ is compared to the doubled nuclear $\mu d + d$ cross section
which represents the first-order approximation to the free molecular
D$_2$ cross section \cite{bracc90,bracc89b}.

Based on the new rate calculations, a Monte Carlo simulation
predicting the time spectra of fusion products from an infinite solid
D$_2$ lattice has been performed. The resulting spectra were fit with
the same kinetics function as the real data to extract effective
rates. The analysis of the Monte Carlo spectra will be presented after
discussing fits to the data.

The suppression of the inelastic scattering probability in solid
deuterium implies a~slow thermalization rate for any epithermal $\mu
d$ atoms, which in turn means that the energy distribution of the
muonic atoms can {\it not}\/ be assumed to be constant in time.  The
implications of these effects on the use of the two-node
approximation are given in the discussion section.

\section{Experimental Equipment}

A description of the cryogenic target system used in this measurement
has been presented elsewhere \cite{knowl96,knowl93}. In brief, a
continuous-flow liquid helium cryostat was used to cool two $51\:\mu
\mathrm{m}$ thick gold foils to 3~K (Fig.~\ref{fig:detplan}). A gas
spraying system directed the deuterium toward the adjacent surfaces of
the foils where it condensed and froze, creating a solid deuterium
film maintained in a high vacuum.  A palladium filter, which operated
above 600~K, was used to purify the D$_2$ immediately prior to
freezing so the ratio of even to odd molecular spins was expected to
be statistical (nD$_2$)\cite{souer}. Muons entered the deuterium after
passing through one thin gold foil, which acted as a degrader, and
stopped either in the first or second solid hydrogen layer, or in the
gold itself.  Muon catalyzed fusion of $d\mu d$ molecules then
occurred. The higher energy fusion products, neutrons at 2.45~MeV and
protons at 3.02~MeV, were detected.

Measurements with solid layers of D$_{2}$ and $^{1}\mathrm H_{2}$ were
made.  Two different layer thicknesses of solid deuterium,
$10.54(53)\:\mathrm{mg}\cdot \mathrm{cm}^{-2}$ (herein called thick)
and $2.11(11)\:\mathrm{mg}\cdot \mathrm{cm}^{-2}$ (thin), $\sim
40\:\mathrm{cm}^{2}$ in area, were frozen onto the foils
\cite{mcfnim}. The thick D$_2$ layer received an integrated muon flux
of $220\times10^{6}\:\mu$ with 53(3)\% stopped in the solid layer,
while the thin layer received $164\times10^{6}\:\mu$ with a stopping
fraction of 14(2)\%.  The study of the background in $^{1}\mathrm
H_{2}$ (where there is no strong fusion process) for both thicknesses
above and for bare target conditions was done with
$130\times10^{6}\:\mu$. Table~\ref{tab:solidd2} summarizes the data.

\subsection{The Trigger}
\begin{figure}
\centerline{\psfig{figure=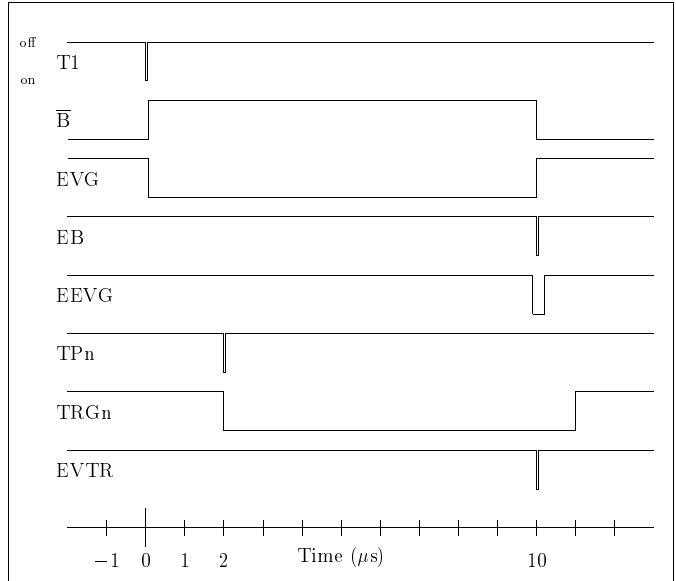,width=8.89cm}}
\caption{The timing diagram for the trigger electronics.
Time is indicated horizontally, and NIM logic level {\it on/off},
indicated by the lower/higher lines. The relationship between the
various signals is explained in the text.}
\label{fig:electim}
\end{figure}

The timing diagram for the trigger electronics is shown in
Fig.~\ref{fig:electim}.  When the T1 scintillator (see
Fig.~\ref{fig:detplan}) signaled the entrance of a muon, two gates
were opened: $\mathrm{B}$ to indicate that a muon was in the target,
and EVG, the event gate.  If any of the detectors recorded an event at
a time when EVG was true, such as the pictured TPn, the corresponding
TRGn gate was opened. At the end of the event gate duration, the
logical OR of all possible TRGn gates was placed in coincidence with
the end of the event gate, EEVG, to determine whether the event was
read (an event trigger) or cleared.  Events for which a second muon
arrived in the target within $\pm 10\:\mu\mathrm{s}$ were discarded
either by inhibiting the trigger or during analysis based on an event
flag.  An event trigger immediately initiated a hardware inhibit to
prevent loss of the data already measured. A computer-driven signal
maintained the inhibit while the data were read and while the CAMAC
modules were cleared. An event clear would inhibit the trigger system
and then remove any values accidently recorded by the CAMAC system.

\subsection{Neutron Detectors}
\begin{figure}
\centerline{\psfig{figure=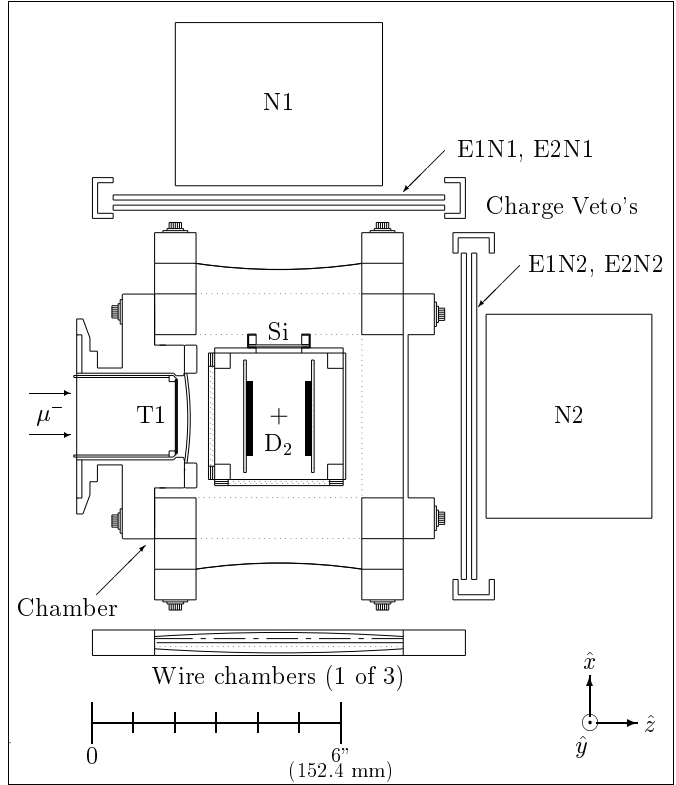,width=8.89cm}}
\caption{A top view of the detector positions with respect to the
target.  Not shown are the remaining two wire chambers, scintillators,
and Sodium Iodide crystal which constituted the imaging system, nor
the Germanium detector, which, when in use, replaced neutron detector
N1.}
\label{fig:detplan}
\end{figure}

The arrangement of detectors is shown in Fig.~\ref{fig:detplan}. Each
neutron detector was a 12.5~cm diameter by 10~cm deep cylinder of
NE--213 liquid scintillator viewed by a photomultiplier.
Hardware-based pulse shape discrimination was used to differentiate
between photon and neutron events. Uncharged particles were selected
by an anticoincidence with both members of a pair of charged-particle
scintillators [EiN1 and EiN2 for $(\mathrm{i}=1,2)$ in
Fig.~\ref{fig:detplan}]. Since photon events were very common,
prescaling was done to keep the photon triggers to an acceptable
level. Typical neutron rates during the measurements were less than
150~s$^{-1}$.

An effective method for background suppression in $\mu{\mathrm{CF}}$
processes is the requirement that the muon survive the event, a
condition enforced by the detection of the muon decay electron.  This
delayed electron, or {\it del\_e}, condition was satisfied if the muon
decay electron was detected between 0.2~and $5.05\:\mu \mathrm{s}$
after the time of a candidate fusion event. It strongly enhanced the
signal-to-background ratio of the data.  Electrons from muon decay
were detected by the coincidence of either pair of charge-veto
scintillators. 

The time spectrum of the decay electrons was used to determine the
proportion of muons stopped in hydrogen. Decay electrons were generated
either with a free muon lifetime (2.2~$\mu \mathrm{s}$), or the
lifetime of muons in Au (79~ns). The intensity of each lifetime
component was proportional to the number of muons stopped in the
respective material.

\subsection{Silicon Charged-Particle Detectors}
A fully depleted, ion implanted silicon detector of
$600\:\mathrm{mm}^{2}$ active area was mounted inside the vacuum
system to detect the charged fusion products. Its $150\:\mu
\mathrm{m}$ active thickness was sufficient to measure protons of up
to 4~MeV, and was well suited for detecting the fusion protons at (or
below) 3.02~MeV. The detector was mounted on the cryostat cold shield,
in front of the neutron detector N1, and viewed the solid deuterium
layers directly (see~Fig.~\ref{fig:detplan}). At the nominal operating
voltage of 30~V the best energy resolution achieved was 170~keV~FWHM
due to mounting constraints and the resulting long cable to the
preamplifier. The 3~MeV protons produced by $dd$ fusion were easily
seen but the tritons and $^3\mathrm{He}$ were too low in energy and
hence obscured in background or stopped in the solid target before
reaching the detector. The rate from silicon detector events was about
25~s$^{-1}$.

\section{Analysis}

\subsection{Neutron Spectra}
Three main event types occur in the neutron spectrum: fusion neutrons,
capture neutrons, and neutrons from the ambient background (zone
neutrons). Fusion neutrons, which constitute the signal, are generated
with a time distribution $n_{f}$, according to the kinetics of the
muon catalyzed fusion reactions in the target and are thus time
correlated with the muon arrival and will satisfy the {\it del\_e\/}
condition with the efficiency $\epsilon_{e}$ with which the electron
detectors measure that a muon has decayed in the delayed time interval
0.2--5.05~$\mu\mathrm{s}$. This efficiency is comprised from a solid
angle contribution and the time window efficiency:
\begin{equation}
\epsilon_{e} = \Omega_{e}\lambda_{\circ}\int^{5.05\;\mu \mathrm{s}}
_{0.2\;\mu \mathrm{s}}
\kern-5mm d\tau \/ e^{-\lambda_{\circ}\tau}.
\end{equation}

Neutrons from nuclear muon capture, with time distribution $n_{c}$,
are also time correlated with the arrival of the muon according to the
capture and decay lifetimes, but are obviously not followed by a muon
decay electron. Only those capture neutrons that are followed by some
accidental event in the electron detectors will be admitted to the
{\it del\_e\/} spectrum (with efficiency $\epsilon_{a}$).

The neutron events from the ambient flux contribute a constant
background, $n_{z}$. These events pass the {\it del\_e\/} cut
depending on the time $t$ of the neutron with respect to the muon
arrival $t_0$.  The {\it del\_e\/} condition requires an event in the
electron detectors between 0.2~$\mu \mathrm{s}$ to 5.05~$\mu
\mathrm{s}$ after the neutron.  For a neutron that occurs more than
5.05~$\mu \mathrm{s}$ before the muon arrival, only an accidental {\it
del\_e\/} condition with efficiency $\epsilon_{a}$ can pass the event
into the {\it del\_e\/} spectrum.  As the time of the ambient
background neutron event moves closer to the subsequent arrival of the
muon, more of the cut interval overlaps with the time when a muon is
known to be in the target and hence the efficiency for accepting the
neutron grows. For a neutron event occurring 0.2~$\mu \mathrm{s}$
before the muon arrival, the time selection interval overlaps exactly
with the muon arrival, hence the muon decay selects the background
event with an efficiency slightly larger than the decay electron
efficiency $\epsilon_{e}$.  After the muon arrival, accidental events
are accepted with the electron detection solid angle efficiency
weighted by the exponential probability that a muon had survived to
the time of neutron detection.  The flat background, $n_{z}$, thus
acquires a muon lifetime dependence in the delayed spectrum. This
relationship is represented by the integral term in
Eq.~(\ref{equ:delens}). The efficiency $\epsilon_{a}$ appears in all
terms since an accidental {\it del\_e\/} condition permits neutrons
into the {\it del\_e\/} spectrum regardless of the neutron source.

The raw (without the {\it del\_e\/} condition) and {\it del\_e\/}
spectra are given by:
\begin{equation}
n_{r}=n_{f}+n_{c}+n_{z}
\label{equ:signs}
\end{equation}
and
\begin{equation}
{\displaystyle n_{del\_e}=\left(\epsilon_{e}+\epsilon_{a}\right)
 n_{f}+\epsilon_{a} n_{c}\qquad\qquad\qquad\qquad\qquad}
\atop{\hfill \displaystyle \mbox{}+
\left(\epsilon_{a}+\Omega_{e}\lambda_{\circ}\!\!
\int^{t+5.05\;\mu \mathrm{s}}_{t+0.2\;\mu \mathrm{s}} 
\kern-5mm d\tau \/
\Theta (\tau-t_0)e^{-\lambda_{\circ}(\tau-t_0)} \right) n_{z},}
\label{equ:delens}
\end{equation}
where $\Theta (\tau-t_0)$ is a step function which models the discrete
arrival time of the muon.

Combining (\ref{equ:signs}) and (\ref{equ:delens}) gives:
\begin{equation}
{\displaystyle n_{del\_e}=\epsilon_{e} n_{f}+ \epsilon_{a} n_{r}
\qquad\qquad\qquad\qquad\qquad\qquad\qquad}\atop
{\displaystyle \mbox{}+ \left(\Omega_{e}\lambda_{\circ}
  \int^{t+5.05\;\mu \mathrm{s}}_{t+0.2\;\mu \mathrm{s}}
  \kern-5mm  d\tau \, 
 \Theta (\tau-t_0)e^{-\lambda_{\circ}(\tau-t_0)} \right)n_{z}\: .}
\label{equ:delefit}
\end{equation}
Note that knowledge of the unknown capture time spectrum, $n_c$, is
not required.  The largest contribution to the background is from the
accidental {\it del\_e\/} coincidences, and is proportional to the raw
distribution $n_{r}$, a spectrum well measured during the experiment
and having small uncertainties. The events in the {\it del\_e\/}
spectrum occurring in the short time interval measured before the muon
arrival determine the value of $n_{z}$.

In Eq.~(\ref{equ:delefit}) there are two input spectra, each with an
uncertainty.  Replacing the integral by the symbol $I(t)$, we have
\begin{equation}
n_{del\_e}\pm \delta n_{del\_e} =
 \epsilon_{e}n_{f}+ \epsilon_{a}\left( n_{r}\pm \delta n_{r} \right)+
 I(t)n_{z}.
\label{equ:nfit}
\end{equation}
The efficiency factor $\epsilon_{e}$ will eventually be absorbed in a
fitted normalization constant, so it does not need to be determined
independently.

\begin{figure}
\centerline{\psfig{figure=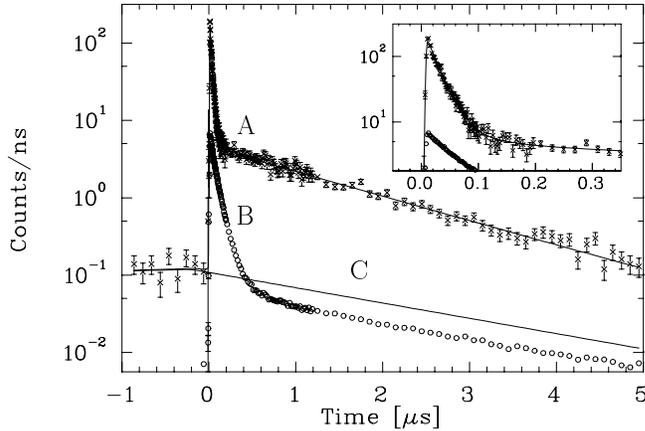,width=8.89cm}}
\caption {Neutron time data assiciated with
Eq.~\protect\ref{equ:delefit}. Data points (A) are the {\it del\_e\/}
spectrum. Points (B)(circle) show the $\epsilon_{a}$-scaled
contribution of the raw spectrum. Curve (C)(line) is the
representation of the other backgrounds, fit assuming a constant value
before time zero, and a muon lifetime after time zero. The fitted
contribution of the fusion neutrons ($n_f$) is shown as the solid line
passing among points (A).}
\label{fig:neutfit}
\end{figure}

The weights for the $\chi^2$ fit depend on the input spectra
uncertainties;
\begin{equation}
{\mathrm{weight}} = \left[ \left( \delta n_{del\_e} \right) ^2
   +   \left( \epsilon_{a}\delta n_{r} \right) ^2 \right]^{-1}.
\end{equation}
The weights contain the variable $\epsilon_{a}$, so the fitting
routine was iterated, each time re-evaluating the weights, until a
stable value for $\epsilon_{a}$ was found.  Figure~\ref{fig:neutfit}
shows the data and the fitted curve.

\subsection{Proton Spectra}
The proton spectra exhibited many features, and careful study of five
different types of target was necessary to extract meaningful
parameters. The different target conditions were: bare target, thick
and thin solid {\em protium}\/ layers upstream and downstream, and
thick and thin solid {\em deuterium}\/ layers upstream and downstream
(see~Table~\ref{tab:solidd2}).

\subsubsection{Energy Spectra}
\begin{figure}
\centerline{\psfig{figure=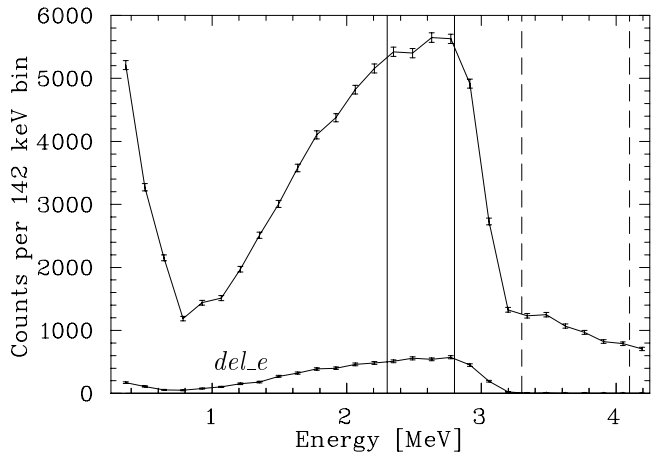,width=8.89cm}}
\caption{A plot of the silicon energy spectrum for solid deuterium.
The 3~MeV proton peak is spread due to energy loss in the thick
target. Note the strong suppression of the background in the {\it del\_e\/}
spectrum. The vertical lines show the $h$~(dashed) and $l$~(solid) energy
cut regions.}
\label{fig:siedf}
\end{figure}

The fusion protons are produced at 3.02~MeV, but lose energy in the
target resulting in a broad distribution: Fig.~\ref{fig:siedf} shows
the spectrum taken with the thick deuterium target.  The background
comprises about 20\% of the total counts.

The background in the detector comes mainly from charged particle
emission following nuclear muon capture.  Protons are the most
commonly emitted particle with a spectrum which is suppressed for
energies below the nuclear Coulomb barrier.  The emission probability
ranges from 15\% for muon capture on silicon---with a 5~MeV Coulomb
barrier---down to about 1\% following muon capture on gold where the
Coulomb barrier is about 16~MeV \cite{lifsh80}.  The relative
intensity of the background depends on the thickness of the hydrogen
layers and the incident muon momentum which together determine the
ratio of muons stopped in the gold target support foil or detector
silicon to those stopped in the hydrogen target itself.  Thick
hydrogen targets exhibited less background.

The background in the silicon detector was removed to the
$\epsilon_{a}$ level by the {\it del\_e\/} requirement, evidence that the
background was due to a process which consumed the muon. The
$\epsilon_{a}$ measured by the ratio of the number of counts above the
3~MeV fusion limit (hence nonfusion related), to those from the same
spectrum after applying the {\it del\_e\/} cut gave a value of 0.11(4)\%,
which is in agreement with the value found for $\epsilon_{a}$ from the
neutron detector: 0.14(4)\%.

\subsubsection{Time Spectra}
Time spectra for each of the five target types were collected for
23~different energy windows, each 142~keV wide (the intrinsic energy
resolution was 170~keV), beginning at 0.9~MeV\@. They were examined
for number, intensity, and value of visible lifetimes, and time-zero
value.  For data where fusion events contributed to the spectrum the
characteristic two-lifetime form of the kinetics in pure deuterium
allowed an accurate measure of $t_0$ for that energy window, as well
as the intrinsic time resolution of the detector at that energy.

The silicon detector exhibited time walk. Thick deuterium data (see
Table~\ref{tab:solidd2}) were chosen to measure the energy dependent
correction since the fusion protons provided a well defined time zero,
and the energy loss in the layer distributed the signal over the
energy range 1--3~MeV\@. Walk gave a 30~ns correction between the
3~MeV and the 1~MeV time spectra, where the difference in proton
flight times at the different energies was no more than 4~ns.

Two standard energy regions were defined from which the time spectra
were selected for further analysis.  The lower energy signal region,
called $l$, was limited to 2.3--2.7~MeV; below 2.3~MeV the time
resolution was poor enough that it began to obscure the structure of
the fast lifetime component in the fusion signal. The limit at 2.7~MeV
was selected to favor fusions occurring away from the surface of the
layer, where the possibility of $\mu d$ escape from the layer would
not affect the kinetics representation. The higher energy background
region, called $h$, was selected between 3.3--4.0~MeV, where there
were no fusion events.  Walk-corrected data were then selected for
the $h$ and $l$ regions.

\begin{figure}
\centerline{\psfig{figure=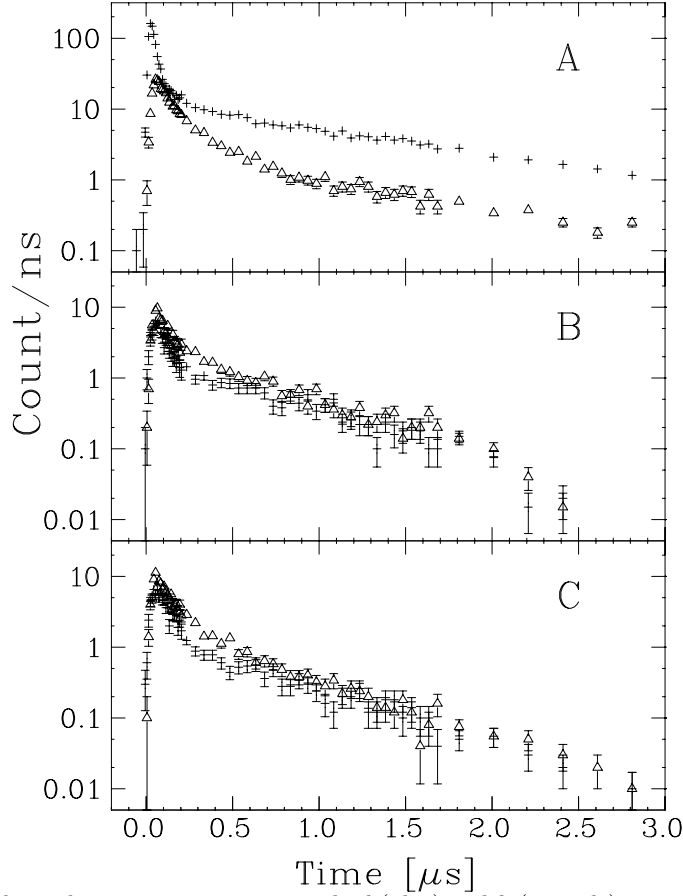,width=8.89cm}}
\caption{Silicon detector time spectra in the $l$~(plus) and
$h$~(triangle) regions in A)~thick solid deuterium, B)~bare target,
and C)~thick protium target.  For B) and C) the $l$~spectrum is
systematically lower than the $h$~spectrum during early times.}
\label{fig:sitdf}
\end{figure}

Figure~\ref{fig:sitdf} shows the time spectra for the $h$~and
$l$~energy cut regions for the thick deuterium, bare target, and thick
protium runs.  Two lifetimes are the dominating feature in the thick
deuterium data, (A), as expected from the two-node approximation. The
fusion spectrum has a rapid onset and rise of the leading edge,
a~shape given by the time resolution of the detector and not by the
physical processes of muon stopping, atomic capture, and
thermalization.

The spectra without fusion signals, the $h$~and $l$~regions for both
gold and thick protium and the $h$~region for thick deuterium, seem
delayed with respect to the fusion signal. It is possible to extract
two lifetimes from the background spectra as well---lifetimes
consistent with muons in gold (80~ns) and muons in intermediate mass
nuclei like silicon or aluminum ($\sim700$~ns). The background does
not pass the {\it del\_e\/} requirement except by $\epsilon_{a}$
coincidence, however the fusion statistics in the {\it del\_e\/} spectrum
were insufficient for the kinetics analysis.

The background in the fusion spectrum was pa\-ramet\-rized and removed
phenom\-eno\-logically. Studies of the $h$~and $l$~time spectra for
targets that did not produce fusion signals provided the method to
predict the background in the $l$~region given the spectrum in the
$h$~region. Scaling from the $h$~region to the $l$~region was time
dependent, {\it i.e.,}
\begin{equation}
l(t) = f(t)\times h(t).
\label{equ:htol}
\end{equation}
This is consistent with a background of charged particles emitted
after muon capture on differing nuclei. Each capture spectrum
contributes to the overall time spectrum with an intensity dependent
on the chosen energy range.  Since the energy spectra of the emissions
are nuclei dependent, the overall time spectrum cannot be expected to
be independent of energy.  The function $f(t)\pm\delta f$,
parametrized in four variables $A_i$ via
\begin{equation}
 f(t)=\left[ A_{1} + A_{2}e^{-A_{3}\left( t-A_{4} \right)} 
\left( 1-e^{-A_{3}\left( t-A_{4} \right)}\right) \right],
\end{equation}
was found using Eq.~(\ref{equ:htol}) to fit the $h$~and $l$~time
spectra from both thin and thick protium layers; $\delta f$ was found
from the covariance matrix of the $A_i$.

Since the {\it del\_e\/} condition was not imposed on the fit data,
the $l$~spectra containing fusion signals were simply composed of the
fusion protons and the background.  The fit expression, with
uncertainties, was
\begin{equation}
p_{l}\pm \delta p_{l} = p_{f} + 
\left( f \pm \delta f \right) \left( p_{h}\pm \delta p_{h} \right) 
\end{equation}
where $p_{l}$ was the measured time spectrum in the $l$~energy window,
$p_{h}$ was from the $h$~energy window, $p_{f}$ was the kinetics
function, after convolution with the detector resolution, for proton
production by $\mu{\mathrm{CF}}$, and $f$ was the background
transformation function as above.

The weight for each fitted point was constructed in the same 
manner as in the neutron case, yielding:
\begin{equation}
{\mathrm {weight}}=\left[ \left(\delta p_{l}\right)^2 +
  \left( f \delta p_{h} \right)^2 +
  \left( p_{h} \delta f \right)^2 \right] ^{-1}.
\label{equ:errpfit}
\end{equation}

\subsection{Fit Methods}
The data were fit with a function made by convoluting the theoretically
expected time distributions, Eq.~(\ref{equ:timestruct}) of the
Appendix, with a Gaussian detector time resolution and a step function
modeling the muon arrival. The time spectra data were rebinned with an
adaptive step size which preserved the fine binning where sensitivity
to fast rates was required, but used larger steps at long times where
small scale sensitivity was not needed. The larger bins at long times
eliminated the problem of fitting data containing many zeros, and
reduced the total number of points in the fit, which in turn
significantly reduced the amount of computer time required per fit.

Excluding background parametrization, six independent parameters were
required to completely determine the shape of a spectrum: two
lifetimes, two amplitudes, the detector timing resolution $\sigma$,
and the muon arrival time $t_0$. There are ten kinetic parameters
which determine the values of the lifetimes and amplitudes, and so it
is impossible to measure more than four of them at any one time.
Since an absolute measurement was not done, one of the parameters was
used for normalization, so only three of the ten kinetics parameters
could be extracted from a single time spectrum. By fitting both
neutron and proton data simultaneously, four kinetics parameters could
be extracted (equivalent to an absolute measurement for a single
fusion product spectrum).  The parametrization of the background in
the data introduced other parameters which were independent of the
kinetics. The fit to the data was made by $\chi^2$ minimization.

The fits to the spectra were sensitive to the values of
$\tilde{\lambda}_{\frac{3}{2}\frac{1}{2}}$,
$\tilde{\lambda}_{\frac{3}{2}}$, $\phi_{z}\lambda_{z}$ (the loss rate
to target contaminants heavier than hydrogen, predominantly nitrogen),
$\tilde{\lambda}_{\frac{1}{2}}$, $\mathrm{P}_{s}$, $\beta_{s}$, and
$\beta_{p}$ (see Fig.~\ref{fig:muind} and the Appendix for the
definition of the parameters). The standard values of the parameters
passed to the fitting routine are given in Table~\ref{tab:fitpass},
along with the uncertainty on the value used when examining the
systematic uncertainties.

\subsection{Fit Results}
For both neutron and proton data, fits were made to the
individual time spectra. Neutron data were restricted to the thick
deuterium targets only, since insufficient statistics existed for the
thin deuterium layers. For the proton data, both thick and thin
deuterium layers yielded spectra which could be fit.

In the fits to individual spectra, the two kinetics
parameters $\tilde{\lambda}_{\frac{3}{2}\frac{1}{2}}$ and
$\tilde{\lambda}_{\frac{3}{2}}$, as well as the loss rate to
high-Z components, $\phi_{z}\lambda_{z}$, were measured.
When the fits to the combined spectra were done, the
sensitivity to four parameters was used to measure
$\tilde{\lambda}_{\frac{3}{2}\frac{1}{2}}$,
$\tilde{\lambda}_{\frac{3}{2}}$, $\phi_{z}\lambda_{z}$, and
values for $\tilde{\lambda}_{\frac{1}{2}}$,
$\mathrm{P}_{s}$, $\beta_{s}$, and $\beta_{p}$, each taken
in turn with the remaining three of the four fixed at the
standard values (see~Table~\ref{tab:fitpass}).

\subsubsection{Uncertainties}
The systematic uncertainties fall into two categories; the effects on
the measured values resulting from uncertainties in input parameters,
and the effects due to variations in the fit interval, cuts on the
input spectrum, and, in the case of the proton spectrum, the
background scaling function. The uncertainties due to input parameters
in the fit are relatively easy to identify and to understand by
changing the parameter and refitting the data. The systematic effects
due to the cuts on the spectrum and the fitting interval were studied
by fitting the same data with different cuts. The fit was considered
satisfactory when the variations on the fit parameters with cut value
were small with respect to the other systematic and statistical
uncertainties.

Within the stated uncertainty, the values found for the fitted
parameters were not affected by variations in temperature T (used to
make $\tilde{\lambda}_{\frac{1}{2}\frac{3}{2}}$ from the
$\tilde{\lambda}_{\frac{3}{2}\frac{1}{2}}$ rate via detailed balance),
density $\phi$, muon decay rate $\lambda_{\circ}$, the detection
efficiency $\epsilon$ (which was detector dependent), and
${\omega}_{\mathrm s}$.  This was checked by changing the input values
of the above parameters by twice their respective uncertainties and
confirming no variation in the fitted parameters. The uncertainty in
the input values of $\tilde{\lambda}_{\frac{1}{2}}$, $\beta_{s}$,
$\beta_{p}$, and $\mathrm{P}_{s}$ did have significant systematic
effects.

The total systematic uncertainty on a fitted parameter was determined
by the addition in quadrature of the uncertainties due to the possible
variations of the fixed input parameters. The systematic uncertainty
associated with the proton background scaling function was handled in
a different way. The uncertainty in the background scaling function
was explicitly taken into account in the fitting function, and so the
uncertainty due to the scaling is included in the statistical error
evaluation (see \S\ref{sec:systemf} below).  Asymmetric uncertainties
were determined by the change necessary to increase the $\chi^2$ of
the fit by 1.0, however, in no case were the uncertainties strongly
asymmetric.

\subsubsection{Neutron Data Fits}
Fits to the thick deuterium neutron data are summarized in
Table~\ref{tab:neutfit} and plotted in Fig.~\ref{fig:neutfit}. The
fitted value for the $\epsilon_{a}$ efficiency agrees with values
measured from neutrons above the fusion energy edge, measurements with
the protium targets, and with measurements made with the silicon
detector.  The $\sigma$ value for the time resolution of the detector
yields a FWHM of 3.5~ns, in accord with the flight time of 2.45~MeV
neutrons traversing a 10~cm deep detector.

\subsubsection{Proton Data Fits} \label{sec:systemf}
\begin{figure}
\centerline{\psfig{figure=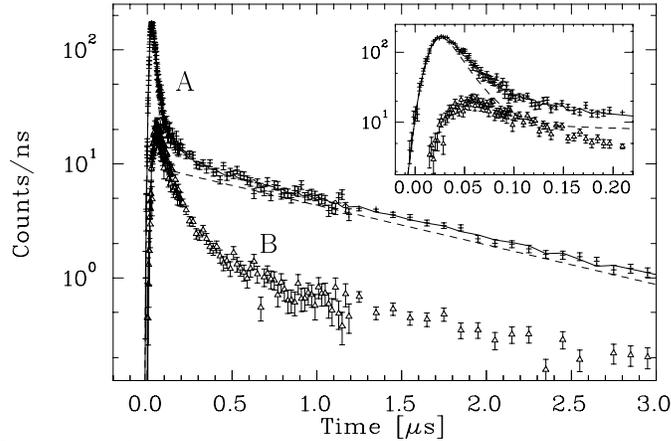,width=8.89cm}}
\caption{Curve (A)(plus) are the proton fusion data measured in the
$l$~energy region. Curve (B)(triangle) shows the background as
determined from the $h$~energy region and the scaling function
$f(t)$. The solid line is the fit and the dashed line shows the
contribution of the fusion kinetics.}
\label{fig:protfit}
\end{figure}

Fits to the proton data taken from both the thick and thin deuterium
layers are summarized in Table~\ref{tab:neutfit}; the results of the
fit for the thick target data are plotted in Fig.~\ref{fig:protfit}.
The results for the thick deuterium given in the table are in
reasonable agreement with the neutron results in
Table~\ref{tab:neutfit}. The signal-to-background ratio for the
thick deuterium proton data was roughly 9:2, while the ratio for the
thin deuterium data was three times worse at 9:6. The background
scaling function was predicted using both the thick and thin protium
layers and so, to check the systematic effects on the fits, the two
deuterium spectra were each fitted with both of the scaling functions.
The results of the fits showed that the thick deuterium data were not
sensitive to the choice of background scaling, presumably due to the
favorable signal-to-background ratio, while the thin deuterium fits
fluctuated to the limits of the statistical accuracy. The data
extracted from the thin deuterium targets were more sensitive to other
effects such as $\mu d$ escape from the layer and, from the value of
$\phi_{z}\lambda_{z}$, suffered from a higher contamination.

\subsubsection{Simultaneous Fits to Both n and p Spectra}
The simultaneous fit of both the neutron and proton data from the
thick deuterium target allowed one more kinetic parameter to be
measured. The choice was limited to $\tilde{\lambda}_{\frac{1}{2}}$,
$\beta_{s}$, $\beta_{p}$, and $\mathrm{P}_{s}$: the kinetics equations
were not sufficiently sensitive to other parameters.  All of the fits
gave a $\chi^{2}$ value of 325 for 309 degrees of freedom.

With $\tilde{\lambda}_{\frac{1}{2}}$ and any two of $\mathrm{P}_{s}$,
$\beta_{s}$, and $\beta_{p}$ fixed, a value for the remaining
parameter was found. The results of the fit, which yielded values for
$\mathrm{P}_{s}$, $\beta_{s}$, and $\beta_{p}$, in addition to
$\tilde{\lambda}_{\frac{3}{2}\frac{1}{2}}$,
$\tilde{\lambda}_{\frac{3}{2}}$, and $\phi_{z}\lambda_{z}$, are given
in columns 2--4 of Table~\ref{tab:allfit}.

The values of $\tilde{\lambda}_{\frac{3}{2}\frac{1}{2}}$,
$\tilde{\lambda}_{\frac{3}{2}}$, and $\phi_{z}\lambda_{z}$ do not
change during the fits to $\mathrm{P}_{s}$, $\beta_{s}$, and
$\beta_{p}$, behavior which is expected from Eq.~(\ref{equ:br}) which
enters the kinetics as a distinct subformula. The large systematic
uncertainty in the $\tilde{\lambda}_{\frac{3}{2}}$ value is due almost
entirely to the uncertainty in the $\tilde{\lambda}_{\frac{1}{2}}$
input value.

Fixing $\mathrm{P}_{s}$, $\beta_{s}$, $\beta_{p}$ at the standard
values allowed a measurement of $\tilde{\lambda}_{\frac{1}{2}}$,
presented in the last column of Table~\ref{tab:allfit}. For this fit,
systematic uncertainty in the parameters is dominated by the
uncertainty in the $\beta_{p}$ input value.

\subsection{Fits to the Monte Carlo Simulations}
A Monte Carlo (MC) simulation has been performed using the
calculated~(Ref.~\cite{adamcDUBNA}) energy-dependent rates for $\mu d$
atom scattering on solid deuterium at 3~K, calculated rates of $d\mu
d$ molecule formation (resonant and nonresonant) on a~free D$_2$
molecule and back-decay rates of the $(d\mu d)d$ complex
\cite{faifm89b,faifmDUBNA}. Since the solid target used was
a~statistical mixture of ortho- and para-deuterium (nD$_2$), the
$\mu d$ scattering on D$_2$ molecules in both $K=0$ and $K=1$ initial
rotational states has been taken into account. The muon transfer from
$\mu d$ atoms to high-Z contamination has been described by a~constant
muon transfer rate.

Recent experiments~\cite{abbotsub,hartmPSI} and theory~\cite{marku94}
show that the initial distribution of $\mu d$ energy contains an
admixture of nonthermalized (in the region of 1--10~eV) $\mu d$ atoms.
The initial $\mu d$ energy in our MC has been described by a~sum of
two Maxwell distributions. One of them, corresponding to atoms
thermalized at 3~K, has a mean energy of 0.4~meV. The hot $\mu d$
atoms are assumed to have a mean energy of 1~eV.  The details of the
shape of the high-energy component are not important for this
simulation because the deceleration rate at energies greater than
about 0.1~eV is very high. Therefore, the main input parameter to the
Monte Carlo is the relative intensity of the energetic $\mu d$
population. The second important parameter is a~scaling factor for the
energy-dependent nonresonant spin flip rate
$\lambda_{\frac{3}{2}\frac{1}{2}}(E)$, to account for the disagreement
between theoretical results and experiment (see~\cite{scrin93}).

A~qualitative agreement with the two-node kinetics model has been
reached by a proper choice of the two important parameters of the
Monte Carlo. Fitted MC spectra yield values of
$\tilde{\lambda}_{\frac{3}{2}}$ and
$\tilde{\lambda}_{\frac{3}{2}\frac{1}{2}}$ consistent with experiment
if the scaling factor for the theoretical nonresonant spin flip rate
is about 0.5 and the relative population of the hot $\mu d$'s is near
0.75.  However, this fit of the kinetics function to the early-time
part of the neutron and proton MC spectra does not yield satisfactory
values of $\chi^2$.  This result is probably due in part to the lack
of a correct form of the energy-dependent $d\mu d$ resonant molecular
formation rate below about 10~meV, in solid deuterium at 3~K and zero
pressure.

\begin{figure}
\centerline{\psfig{figure=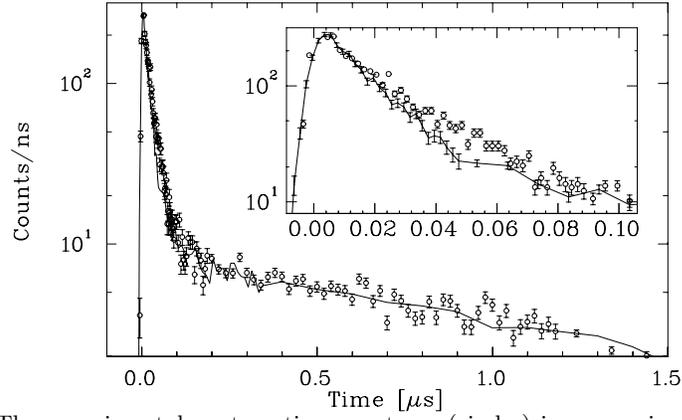,angle=90,width=8.89cm}}
\caption{The experimental neutron time spectrum (circles) in
comparison with Monte Carlo predicted spectrum (line) scaled by a
single normalizing factor. The inset shows also the uncertainty in the
MC spectrum, illustrating the discrepancy at early times.}
\label{fig:mcspect}
\end{figure}

Figure~\ref{fig:mcspect} shows the experimental neutron data in
comparison to the output of the MC simulation. The spectrum at long
times, when $\mu d$ atoms disappear mainly via nonresonant $d\mu d$
formation and muon transition to the Z$>1$ contaminants, is well
described by the MC simulations and good values of $\chi^2$ are
obtained ($\chi^2=33.5/30$~points). This is so since the rates for
nonresonant $d\mu d$ formation, in both the $\mathrm{J}=0$ and
$\mathrm{J}=1$ states, are not strongly energy dependent below 10~meV,
meaning that the detailed shape of the $\mu d$ energy distribution is
not significant. The disagreement at times dominated by the quartet to
doublet hyperfine transition and thermalization is clearly visible as
the steeper slope in the MC spectrum.

\section{Discussion and Conclusions}

Results derived from our experiment show that the processes governing
molecular formation in the solid state are not yet quantitatively
understood, but qualitative consistency has been achieved with the
theory of slow thermalization caused by solid state effects
\cite{adamcDUBNA}.

\subsubsection{Kinetics Parameters}
The values presented in Table~\ref{tab:allfit} for the kinetics
parameters (under the assumptions inherent in the two-node
approximation) represent the first examination of the muon catalyzed
fusion process in solid D$_2$.

The qualitative success of the Monte Carlo simulation of fusion in
solid D$_2$ is a clear indication that slow thermalization effects are
important. It is surprising that the two-node approximation works at
all in this situation. Why it does work can be seen by examining the
processes themselves.

For $\mu d_{\frac{3}{2}}$, slow thermalization at the lowest energies
allows molecular formation via the lowest resonances. Even so, spin
deexcitation processes and thermalization remove $\mu d$ from the
resonant energy region at least ten times faster than the effective
formation rate. Under such conditions, sensitivity to the shape of the
resonance is suppressed and use of a constant rate is adequate.

Other explanations for the large $\tilde{\lambda}_{\frac{3}{2}}$ rate
again rely on the solid state of the deuterium.  Examination of
Eq.~(\ref{equ:effhypform}) shows that an increase in
$\tilde{\lambda}_{f}$, the deexcitation rate leading to fusion for the
resonantly formed $\left[(d \mu d)_{J\nu}dee\right]$, will increase
the effective formation rate by more successfully competing with the
$\Gamma_{SF'}$ deexcitation (back decay). If there exists a
sufficiently strong coupling of the rovibrational modes of the
resonant six-body complex to phonons in the solid, then there may be
sufficient augmentation of the deexcitation rate to increase
$\tilde{\lambda}_{f}$.  The molecular formation process occurring in
solids has not yet been rigorously examined.

Since resonance formation involving $\mu d_{\frac{1}{2}}$ occurs at
collision energies greater than 50~meV, the rapid initial cooling of
the $\mu d_{\frac{1}{2}}$ permits only nonresonant molecular
formation, thus assuming a time-independent nonresonant
$\tilde{\lambda}_{\frac{1}{2}}$ rate is valid.  Our measured value for
$\tilde{\lambda}_{\frac{1}{2}}\;[=0.052(8)_{stat.}(3)_{sys.}\:
\mu\mathrm{s}^{-1}]$ is in agreement with nonresonant molecular
formation rates measured in other experiments \cite{petitDUBNA}.
However, the assumption that $\tilde{\lambda}_{\frac{1}{2}}$ measured
in gas is appropriate for 3~K solid contributes the dominant
systematic uncertainty in our measurement of
$\tilde{\lambda}_{\frac{3}{2}}$.

The $\tilde{\lambda}_{\frac{3}{2}\frac{1}{2}}$ rate measures the
disappearance rate for $\mu d$ atoms capable of participating in
resonant formation via $\tilde{\lambda}_{\frac{3}{2}}$ and thus
contains hyperfine deexcitation from direct scattering, a contribution
from back decay, {\em and} the effective rate of energy loss for the
$\mu d_{\frac{3}{2}}$ which are in the energy window of the lowest
resonance. Since our measured value for the effective hyperfine
deexcitation is consistent with calculations of the rate due to
scattering alone, the calculated rates are too large, a result
consistent with recent $\mu d+\mathrm{HD}$ experiments
\cite{petitDUBNA}.

Emission of the $\mu d$ from the finite solid target layer is a loss
mechanism that is not explicitly accounted for in the analysis,
however the process would mimic the $\phi_{z}\lambda_{z}$ loss
mechanism.  We believe that the contamination was due to nitrogen,
based on the observation of muonic nitrogen x rays.  The value found
for $\phi_{z}\lambda_{z}$, coupled with the value for the transfer to
nitrogen, $\lambda_{z}\approx 10^{5} \:\mu \mathrm{s}^{-1}$, implies a
nitrogen contamination of about 1.5~ppm, a value which was
independently measured for the thick deuterium target via x-ray
analysis. If the value were to represent only $\mu d$ escaping from
the layer, then roughly one third of all the $\mu d$ atoms are emitted
from the layer. Given the current understanding of the scattering
cross sections, this process is unlikely to give such a strong
effect. A Monte Carlo which includes the cross sections corrected for
solid state effects as well as an accurate modeling of the finite
geometry of the targets is required for the quantitative
reinterpretation of the $\tilde{\lambda}_{\frac{3}{2}}$ and
$\tilde{\lambda}_{\frac{3}{2}\frac{1}{2}}$ rates in terms of
thermalization.

Demin {\it et~al.} \cite{deminDUBNA} have presented measured values
for $\mu{\mathrm{CF}}$ rates in solid deuterium over a range of
temperatures, assuming a similar kinetics function. Their results are
consistent with ours, thus the disagreement with theory indicates that
further calculations for the molecular formation rate in solid
hydrogen must be made.  The next experiments will have to examine the
thermalization process, as well as the possible para-ortho
effects. One suggested increase in the molecular formation rate for
thermalized $\mu d_{\frac{3}{2}}$ occurs only for para-D$_2$ ({\it
i.e.,} $J=1$) \cite{menshDUBNA}. Using a solid ortho-D$_2$ target
would help to verify or discount that effect.

In solid hydrogen the molecular formation rate is concentration
dependent, proportional to $\mathrm c_{d}\phi N_{\circ}$, however, the
thermalization rate is proportional to $\phi N_{\circ}$, since the
scattering and energy loss processes for both $\mu d+{\mathrm D}_{2}$
and $\mu d+{}^1\!\mathrm H_{2}$ are similar for low energy $\mu d$.
If slow $\mu d$ thermalization is the process increasing the
$\tilde{\lambda}_{\frac{3}{2}}$ molecular formation rate, an
experiment in a solid mixture of $\mathrm D_{2}+{}^1\!\mathrm H_{2}$
using several values of the deuterium concentration will not yield the
same reduced formation rate $\tilde{\lambda}_{\frac{3}{2}}$.  Such an
experiment in a $\mathrm D_{2}+{}^1\!\mathrm H_{2}$ mixture would be
more difficult to analyze, since the $\mu p$ to $\mu d$ transfer and
subsequent creation of an epithermal $\mu d$ would change the initial
intensity of the epithermal $\mu d$ atoms---the effect believed to be
responsible for the elevated $\tilde{\lambda}_{\frac{3}{2}}$ rate.

\subsubsection{Branching Ratios}
Three values parametrize ${\beta}_{F}$, the effective branching ratio:
$\beta_{s}$, $\beta_{p}$, and $\mathrm{P}_{s}$,
[cf.~Eq.~(\ref{equ:br})].  Table~\ref{tab:allfit} lists fitted values
for each of these parameters determined using standard values for the
other two.

Our value $\beta_{p} = 0.563(14)_{stat.}(11)_{sys.}$ is close to the
0.59 predicted by theory (Hale, \cite{halex90}), and is consistent
with a previous measurement by Balin {\it et~al.} \cite{balin84}.  Our
value, $\beta_{s} = 0.487(15)_{stat.}(11)_{sys.}$, is consistent with
0.47, the prediction of Hale.

Standard values for $\beta_{s}$ and $\beta_{p}$ were used in a fit to
$\mathrm{P}_{s}$ since $\beta_{s}$ and $\beta_{p}$, as discussed
above, were consistent with both theory and experiment.  Faifman
\cite{faifm89b} predicts $\mathrm{P}_{s}=0.560$ at very low energy.
Our result, $\mathrm{P}_{s}=0.47(8)_{stat.}(6)_{sys.}$ is in
agreement.  It should also be noted that the $\mathrm{P}_{s}$ value is
dependent on the energy at which molecular formation occurs, so for
$\mu d$ atoms in the process of thermalizing, a constant
$\mathrm{P}_{s}$ does not accurately represent the true physics
processes.

\section*{Acknowledgments}

This work was supported by the Natural Sciences and Engineering
Research Council (NSERC) of Canada, by a NATO Linkage Grant
(LG~930162), and grants from DOE and NSF in the United States.
A.~Adamczak wishes to thank the Polish Committee for Scientific
Research for support under Grant \#2P03B01809. F.~Mulhauser and
R.~Jacot-Guillarmod thank the Swiss National Science Foundation for
support, and J.M.~Bailey thanks the Leverhulme Trust.

\section*{APPENDIX: Reaction Kinetics}

\begin{figure}
\centerline{\psfig{figure=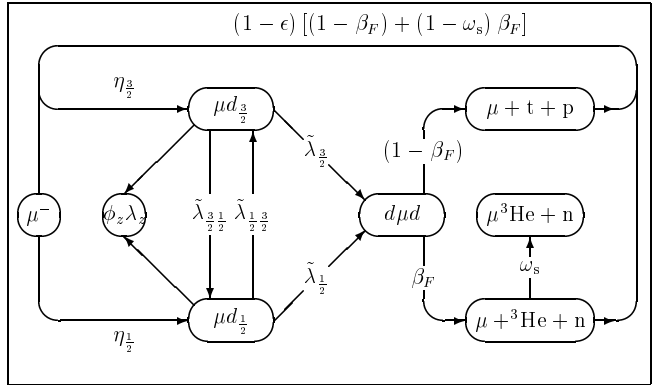,width=8.89cm}}
\caption{The two--node approximation of the deuterium
fusion kinetics. See the Appendix for details and definitions.}
\label{fig:muind}
\end{figure}

This appendix describes in more detail the kinetics equations derived
from Fig.~\ref{fig:muind}.  All rates are normalized to liquid
hydrogen density ($4.25\times10^{22}\:\mathrm{atoms\cdot cm}^{-3}$),
and unit concentration, thus giving reduced rates for comparison with
experiments under different conditions of concentration and
density. The reduced rate of formation of $d\mu d$ molecules from a
$\mu d$ atom in the hyperfine state F is $\tilde{\lambda}_{F}$
\cite{mensh87}:
\begin{equation}
\tilde{\lambda}_{F} = \lambda_{nr} + \sum_{\mathrm S}\lambda_{FS} 
\frac{\tilde{\lambda}_{f}}{\tilde{\lambda}_{f}+\sum_{F''}\Gamma_{SF''}}.
\label{equ:effhypform}
\end{equation}
Molecular formation leading to fusion is composed of the
nonresonant formation rate and the fraction of resonantly
formed molecules which successfully fuse via
$\tilde{\lambda}_{f}$ rather than resonantly scatter as
characterized by $\Gamma_{SF'}$ \cite{mensh87}.
Likewise, the reduced hyperfine transition rate from state F
to $\mathrm{F}'$,\/ $ \tilde{\lambda}_{FF'}$ , has
contributions from regular scattering and from the decay of
the resonantly formed $[(d\mu d)dee]$ complex:
\begin{equation}
\tilde{\lambda}_{FF'} = \lambda_{FF'} + \sum_{\mathrm S}\lambda_{FS}
\frac{\Gamma_{SF'}}{\tilde{\lambda}_{f} + \sum_{F''}\Gamma_{SF''}}.
\label{equ:effspinflip}
\end{equation}
Finally, ${\beta}_{F}$ represents the fusion branching ratio for
$d\mu d$ molecules formed from a $\mu d$
in hyperfine state F\@. Experimentally, ${\beta}_{F}$
is given by
$${\beta}_{F}=\frac{\mu d_F + \mathrm D_{2}
\rightarrow \mathrm{\mu + ^{3}\!He + n} }
{\left( \mu d_F\! +\! \mathrm D_{2}
\!\rightarrow\! \mu + p  + t\right)\! +\! \left(
\mu d_F\! +\! \mathrm D_{2}
\!\rightarrow\! \mathrm{\mu + ^{3}\!He + n} \right) } \/.$$
The $\beta_{F}$ measure will differ between the two hyperfine states
due to the different distributions of bound state angular momenta
selected by the hyperfine-dependent molecular formation processes
\cite{faifm89b}. The ${\beta}_{F}$ parameters are convenient when
writing the kinetics, but are composed of the more fundamental s~and
p~wave fusion branching ratios, together with $\mathrm{P}_{s}$, the
ratio of s~to p~wave bound states from nonresonant formation\cite{zmesk90}. 
Resonance formation always produces the p~wave molecular state. 
\begin{equation}
{\beta}_{F}= \frac{\lambda_{nr}}{\tilde{\lambda}_{F}} 
\left[ {\mathrm{P}_{s}}\beta_{s} + \left(1-{\mathrm{P}_{s}} \right)\beta_{p} \right]
+\frac{\tilde{\lambda}_{F}-\lambda_{nr}}{\tilde{\lambda}_{F}}\beta_{p}.
\label{equ:br}
\end{equation}

The kinetics of the muon catalyzed reactions in pure deuterium are
represented in Fig.~\ref{fig:muind}. Using effective rates, the time
evolution of the two different hyperfine populations can be
approximated as a differential equation of the form:
\begin{equation} 
\frac{d}{dt} {                  \left(
\! \begin{array}{c} N_\frac{3}{2} \\
    N_\frac{1}{2} \end{array}\! \right)}
\! =\! \left( \begin{array}{cc} -A & B \\ 
                                C & -D \end{array} \right) 
\! \cdot \!                       \left(
\! \begin{array}{c} N_\frac{3}{2} \\
    N_\frac{1}{2}  \end{array} \! \right) 
\label{equ:muind}
\end{equation}
with initial conditions $\eta_{F}$ (normalized to a single muon)
$$                                       {\left(
\! \begin{array}{c} N_{\frac{3}{2}}(t=0) \\
 N_{\frac{1}{2}}(t=0)  \end{array}\!      \right)}
\! = \!                                   
\left( 
\begin{array}{c} \eta_{\frac{3}{2}} \\
\eta_{\frac{1}{2}} \end{array}\right)
\qquad \mathrm{where}\quad
\left\{
\begin{array}{c} \eta_{\frac{3}{2}}=\frac{2}{3}, \\
\eta_{\frac{1}{2}}=\frac{1}{3}. \end{array}\right.$$

The coefficients in the matrix can be read from Fig.~\ref{fig:muind}
giving (here all explicitly positive):
$$ {\displaystyle A =  \lambda_{\circ} + \phi_{z}\lambda_{z}
\qquad\qquad\qquad\qquad\qquad\qquad\qquad }\atop{\hfill 
\mbox{}+ \phi  \left\{\tilde{\lambda}_{\frac{3}{2}}
\left[1-\eta_{\frac{3}{2}} \left( 1-\epsilon \right)
     \left( 1-\omega_{\mathrm s}\beta_{\frac{3}{2}}\right) \right] 
   + \tilde{\lambda}_{\frac{3}{2}\frac{1}{2}}      \right\}} $$
$$ B =  \phi \left[ \tilde{\lambda}_{\frac{1}{2}} \eta_{\frac{3}{2}}
     \left( 1-\epsilon \right)
     \left( 1-\omega_{\mathrm s}\beta_{\frac{1}{2}} \right)
   + \tilde{\lambda}_{\frac{1}{2}\frac{3}{2}} \right]\qquad$$
$$ C =  \phi \left[ \tilde{\lambda}_{\frac{3}{2}} \eta_{\frac{1}{2}}
     \left( 1-\epsilon \right)
     \left( 1-\omega_{\mathrm s}\beta_{\frac{3}{2}} \right) 
   + \tilde{\lambda}_{\frac{3}{2}\frac{1}{2}}\right]\qquad$$
$${\displaystyle D =  \lambda_{\circ} + \phi_{z}\lambda_{z}
\qquad\qquad\qquad\qquad\qquad\qquad\qquad }\atop{\hfill 
\mbox{}+\phi \left \{\tilde{\lambda}_{\frac{1}{2}}
\left[1-\eta_{\frac{1}{2}} \left( 1-\epsilon \right)
\left( 1-\omega_{\mathrm s}\beta_{\frac{1}{2}}\right)\right] 
+ \tilde{\lambda}_{\frac{1}{2}\frac{3}{2}} \right\}}.$$

The formal solution of Eq.~(\ref{equ:muind}) to obtain the
time dependence of the hyperfine populations, $N_{F}(t)$, is
tedious but straightforward. Once derived, the time
structure of the fusion products, either protons or neutrons
(generically denoted $k$), can be found by summing over the
populations, formation rates, and branching ratios:
\begin{equation}
 \frac{dk}{dt} = \phi \!\! \sum_{F=\frac{1}{2} , \frac{3}{2} } 
               \!\! \alpha_{F} \tilde{\lambda}_{F} N_{F}
 \quad \mbox{with}\quad\alpha_{F} = \left\{\!\begin{array}{ll}
                  \beta_{F}&\!\mbox{for $k=n$,}\\
              1 - \beta_{F}&\!\mbox{for $k=p$.}   
                      \end{array} \right.
\end{equation}

The time distribution of product $k$ is thus:
\begin{equation}
\frac{dk}{dt}=\Psi_{\frac{3}{2}} e^{L_{\frac{3}{2}} t} + 
              \Psi_{\frac{1}{2}} e^{L_{\frac{1}{2}} t} ,
\label{equ:timestruct}
\end{equation}
with rates which are the negatives of the the parameters:
\begin{equation}
 L_{\frac{3}{2}} = \frac{-1}{2} \left[ (A+D) + \sqrt{(A-D)^{2} + 4BC} \right],
\end{equation}
\begin{equation}
 L_{\frac{1}{2}} = \frac{-1}{2} \left[ (A+D) - \sqrt{(A-D)^{2} + 4BC} \right],
\end{equation}
and amplitudes:
\begin{equation}
\Psi_{\frac{3}{2}}=\frac{{
     {\displaystyle \phi \left \{\alpha_{\frac{3}{2}}
      \tilde{\lambda}_{\frac{3}{2}} \left [ 
      \eta_{\frac{3}{2}} \left ( L_{\frac{3}{2}} + D \right )
    + \eta_{\frac{1}{2}} B \right ]\right.\qquad}\atop{\hfill \displaystyle
      \left. \mbox{}+ \alpha_{\frac{1}{2}} \tilde{\lambda}_{\frac{1}{2}} \left [ 
      \eta_{\frac{3}{2}} C - \eta_{\frac{1}{2}} 
      \left ( L_{\frac{1}{2}} + D \right ) \right ] \right \}} }}
      {L_{\frac{3}{2}} - L_{\frac{1}{2}}}
\label{equ:ampq}
\end{equation}
and,
\begin{equation}
\Psi_{\frac{1}{2}}=\frac{{ 
      {\displaystyle \phi \left \{\alpha_{\frac{3}{2}}
      \tilde{\lambda}_{\frac{3}{2}} \left [
    - \eta_{\frac{3}{2}} \left ( L_{\frac{1}{2}} + D \right )
    - \eta_{\frac{1}{2}} B \right ]\right.\qquad}\atop{\hfill \displaystyle
      \left. \mbox{}+ \alpha_{\frac{1}{2}} \tilde{\lambda}_{\frac{1}{2}} \left [
    - \eta_{\frac{3}{2}} C + \eta_{\frac{1}{2}}
      \left ( L_{\frac{3}{2}} + D \right )  \right ] \right \}} }}
      {L_{\frac{3}{2}} - L_{\frac{1}{2}}}.
\label{equ:ampd}
\end{equation}

The total yield of fusion particles from a single muon can
be obtained by integrating Eq.~(\ref{equ:timestruct}) over
times $[0,\infty)$ to obtain:
\begin{equation}
Y_{k}=-\left( \frac{\Psi_{\frac{3}{2}}}{L_{\frac{3}{2}}} +
              \frac{\Psi_{\frac{1}{2}}}{L_{\frac{1}{2}}} \right) .
\label{equ:yield}
\end{equation}

The above representation of the kinetics is the exact
solution to the two-node approximation used in Zmeskal {\it
et~al.}~\cite{zmesk90}, and verifies the solution presented
therein to within 1\% accuracy.



\begin{table}
\caption{Summary of the data.}
\begin{center}
\begin{tabular}{ccrc}
 Hydrogen  &   Name & \multicolumn{1}{c}{$\mu^{-}$} & \% stopped in \\
\mbox{[mg$\cdot\mathrm{cm}^{-2}\cdot\mathrm{(foil)}^{-1}$]} & & $(\times 10^{6})$ & hydrogen \\ \hline
10.54(53) D$_2$               & Thick Deuterium & 216.611 &53(3)\%   \\
 2.11(11) D$_2$               & Thin Deuterium  & 163.908 &14(2)\%   \\
  0                           & Bare Target     &  18.504 &na     \\
 1.06(6)  $^{1}\mathrm H_{2}$ & Thin Protium    &  48.974 &14(2)\%   \\
 5.27(27) $^{1}\mathrm H_{2}$ & Thick Protium   &  59.797 &53(3)\%   \\
\end{tabular}
\end{center}
\label{tab:solidd2}
\end{table}

\begin{table}
\caption{Standard values used for the fits.}
\begin{center}
\begin{tabular}{ccc}
Parameter  & Value    & Source \\ \hline
${\omega}_{\mathrm s}$     
  & 0.122(3) 
     & Experiment: Balin~{\it et~al.} \cite{balin84b} \\
$\phi$     
  & 1.4269(4)
     & Experiment: R.H.~Sherman in \cite{petit88}  \\
$\tilde{\lambda}_{\frac{1}{2}\frac{3}{2}}$     
  & $0\:\mu \mathrm{s}^{-1}$
     & Detailed balance of $\tilde{\lambda}_{\frac{3}{2}\frac{1}{2}}$ at 3~K \\
$\tilde{\lambda}_{\frac{1}{2}}$ 
  & $0.044(5)\:\mu \mathrm{s}^{-1}$ 
     &  Review: Scrinzi~{\it et~al.} \cite{scrin93} \\
$\mathrm{P}_{s}$  
  & 0.560    
     & Theory: Fa{\u{\i}}fman, \cite{faifm89b} \\ 
$\beta_{s}$
  & 0.47     
     & Theory: Hale, \cite{halex90} \\
$\beta_{p}$
  & 0.580(5) 
     & Experiment: Balin {\it et~al.} \cite{balin84}
\end{tabular}
\end{center}
\label{tab:fitpass}
\end{table}

\begin{table}
\caption{The kinetics values resulting from fits to the individual detector spectra.}
\begin{center}
\begin{tabular}{cccc}
Parameter
     & neutron     & \multicolumn{2}{c}{proton}\\ \hline
     & thick D$_2$ & thick D$_2$ & thin D$_2$\\ \hline
$\tilde{\lambda}_{\frac{3}{2}\frac{1}{2}}\:[\mu \mathrm{s}^{-1}] $ 
     & 35.3(1.4) & 34.7(1.5) & 38.6(3.1) \\
$\tilde{\lambda}_{\frac{3}{2}}\:[\mu \mathrm{s}^{-1}]$
     &3.12(14)&2.77(12)& 2.97(23)\\
$\phi_z\lambda_{z}\:[\mu \mathrm{s}^{-1}]$
     & 0.288(15)  & 0.340(12)  & 0.589(24)\\
$\epsilon_{a}$ & 0.14(4) \% & na & na \\
$\sigma$~[ns]  & 1.48(11)   & 9.52(28) & 9.02(36)\\
$\chi^2$/dof & 158/150 & 164/157 & 162/157 \\
 cl & 55\% & 33\% & 39\% \\
\end{tabular}
\end{center}
\label{tab:neutfit}
\end{table}

\begin{table}
\caption{The kinetics values resulting from a simultaneous fit
to both neutron and proton spectra from the thick deuterium target, see
the text for a complete explanation.
The fixed values are taken from Table~\ref{tab:fitpass}.}
\begin{center}
\begin{tabular}[t]{ccccc}
Parameter       & \multicolumn{4}{c}{Value(Statistical)(Systematic)} \\   \hline
$\tilde{\lambda}_{\frac{3}{2}\frac{1}{2}}\:[\mu\mathrm{s}^{-1}]$ 
&\multicolumn{3}{c}{$34.2(8)(1)$}      & $34.0(8)(1) $ \\
$\tilde{\lambda}_{\frac{3}{2}}\:[\mu\mathrm{s}^{-1}]$
&\multicolumn{3}{c}{$2.71(7)(32)$}     & $3.21(51)(16)$\\
$\phi_{z}\lambda_{z}\:[\mu\mathrm{s}^{-1}]$
&\multicolumn{3}{c}{$0.320(10)(1)$}    & $0.320(10)(1)$\\
$\tilde{\lambda}_{\frac{1}{2}}\:[\mu\mathrm{s}^{-1}]$ 
                 & fixed        & fixed           & fixed           & $0.052(8)(3)$ \\
$\mathrm{P}_{s}$ &{ $0.47(8)(6)$} & fixed           & fixed           & fixed\\
$\beta_{s}$      & fixed        & {$0.487(15)(11)$} & fixed           & fixed\\
$\beta_{p}$      & fixed        & fixed           & {$0.563(14)(11)$} & fixed\\
\end{tabular}
\end{center}
\label{tab:allfit}
\end{table}

\end{document}